\documentclass[
reprint,
nofootinbib,
amsmath,amssymb,amsbsy,
prd,
twocolumn,
superscriptaddress,
longbibliography,
preprintnumbers]{revtex4-1}
\usepackage[dvipsnames,usenames]{xcolor}
\usepackage{graphicx}
\usepackage{subcaption}
\usepackage{dcolumn}
\usepackage{bm}
\usepackage{mathrsfs,natbib}
\usepackage[title]{appendix}
\usepackage{graphicx,epsf,amssymb,amsbsy,amsfonts,amssymb,amsmath}
\usepackage[colorlinks=true]{hyperref}
\hypersetup{
    unicode=false,          
    pdftoolbar=true,        
    pdfmenubar=true,        
    pdffitwindow=false,     
    pdfstartview={FitH},    
    pdftitle={},    
    pdfauthor={},     
    pdfsubject={},   
    pdfcreator={},   
    pdfproducer={}, 
    pdfkeywords={} {} {}, 
    pdfnewwindow=true,      
    colorlinks=true,       
    linkcolor=magenta, 
    citecolor=blue,        %
    filecolor=magenta,      
    urlcolor=blue           
}
\usepackage{slashed}
\usepackage{comment}
\usepackage[dvipsnames]{xcolor}
\usepackage{color,soul}
\usepackage{simplewick}

\newcommand{\be}{\begin{equation}}
\newcommand{\ee}{\end{equation}}
\newcommand\beq{\begin{eqnarray}}
\newcommand\eeq{\end{eqnarray}}

\newcommand{\mybar}[1]

\newlength{\backup}

\begin{document}

\title{Gauge field flow for chiral gauge theories on a slab}
\author{Jinlong Dang}%
\email{dangjl@stu.pku.edu.cn}
\affiliation{School of Physics, Peking University, Beijing 100871, China}
\author{Rohith Karur}%
\email{r\_karur137@berkeley.edu}
\affiliation{Department of Physics, University of California, Berkeley, CA 94720, USA}
\affiliation{Nuclear Science Division, 
Lawrence Berkeley National Laboratory, Berkeley, 
CA 94720, USA}
\author{Srimoyee Sen}%
\email{srimoyee08@gmail.com}
\affiliation{Department of Physics and Astronomy, Iowa State University, Ames, Iowa 50011, USA}%

\date{\today}

\begin{abstract}
The proposal to formulate chiral gauge theories using domain wall fermions on $2n+1$ dimensional Euclidean lattice with a slab geometry \cite{Grabowska:2015qpk} involves $2n$ dimensional dynamical gauge fields residing on one of the domain walls. The gauge fields are extended into the extra dimension using gradient flow decoupling the mirror fermions on the anti-wall. We implement this construction on the lattice for $n=1$ in the presence of $2n$ dimensional background gauge fields. We also formulate and implement an additional gauge field flow proposal \cite{Kaplan:2023pxd, Kaplan:2024ezz}, where the gauge fields satisfy $2n+1$ dimensional equation of motion away from the domain wall, known as the EOM (equation of motion) flow.
In both cases, we couple the gauge fields to fermions and demonstrate how current conservation and anomaly inflow work on the lattice.    
\end{abstract}
\maketitle

\section{Introduction}
Chiral gauge theories (CGT) are notoriously difficult to define on the lattice. This has prevented a complete non-perturbative definition of the standard model which itself is a chiral gauge theory.
Thus, a lattice formulation of chiral gauge theories is highly sought after. There have been several attempts to provide such a formulation in the past \cite{Eichten:1985ft,Kaplan:1992bt,  Luscher:1998du, Wen:2013ppa, Kadoh:2007wz, Lykken:1997gy, Poppitz:2010at}, as well as more recently \cite{Thorngren:2026ydw, You:2014vea, Razamat:2020kyf, Zeng:2022grc}. These include, the Luscher formulation \cite{Luscher:1998du}, Shamir-Golterman approach \cite{PhysRevD.70.094506} and symmetric mass generation \cite{Eichten:1985ft, Zeng:2022grc, Catterall:2020fep, You:2014sqa, You:2014vea, Catterall:2020fep}. The latter was first discussed in \cite{Eichten:1985ft}, and has seen a recent revival \cite{Zeng:2022grc, Wen:2013ppa, Wang:2022ucy}. Other approaches include bosonization \cite{Berkowitz:2023pnz} and the use of symmetry disentanglers \cite{Thorngren:2026ydw}. In this paper, we will focus on what we refer to as the extra dimension based approach or domain wall fermion based approach \cite{Kaplan:1992bt, Grabowska:2015qpk, Kaplan:2023pxd, Kaplan:2023pvd,  Clancy:2024bjb}. This approach was first used to solve the problem of formulating global chiral symmetry on the lattice \cite{Kaplan:1992bt, Shamir:1993zy} and has been extremely successful at simulating quantum chromodynamics (QCD) with good global chiral symmetry \cite{RBC:2014ntl, RBC:2010qam, RBC:2012ynq}. The approach relies on considering a finite $2n+1$ dimensional manifold, $\mathbb{T}^{2n+1}$ with massive Dirac fermions where its mass depends on the extra dimension $x_{2n+1}$: there exists pair of mass defects in $x_{2n+1}$  where the mass changes sign. With such a mass profile for the higher dimensional theory, the two defects are found to host low lying $2n$ dimensional Weyl fermions modes of opposite chirality: e.g. if the wall hosts a positive or right chirality Weyl fermion, the anti-wall hosts a negative or left chirality Weyl mode and vice versa. The modes on the anti-wall are known as mirror fermions. The opposite chirality modes on the wall and the anti-wall can be used to formulate a $2n$ dimensional Dirac fermion with good global chiral symmetry. The fermion sector is then augmented by a $2n$ dimensional gauge field sector where the gauge fields are independent of the extra dimension $x_{2n+1}$ and give rise to a $2n$ dimensional vector-like theory with good global chiral symmetry, e.g. $4$ dimensional QCD for $n=2$.   
A straightforward application of this idea does not work for chiral gauge theories since the fermion sector and the gauge field prescription inevitably result in a vector like theory. In \cite{Grabowska:2015qpk}, Grabowska and Kaplan proposed a modification to this proposal that retains the geometry of the manifold, but proposes to decouple the mirror fermions by using gradient flow for gauge fields in the extra dimension. We refer to this as the slab proposal. A more recent proposal to formulate lattice chiral gauge theories based on the domain wall idea, modifies the manifold to $\mathbb{T}^{(2n-1)}\times \mathbb{R}^2$ and employs a mass defect with a single boundary \cite{Kaplan:2023pxd, Kaplan:2023pvd}, i.e. a disk defect, to engineer an unpaired Weyl fermion on the lattice while gapping out the mirror sector. For related work, see \cite{Aoki:2022aez, Aoki:2022cwg, Kroth:2025hba}. In this disk formulation of CGT \cite{Kaplan:2023pxd, Kaplan:2023pvd} , the gauge fields are extended into the bulk using $2n+1$ dimensional equation of motion. We refer to this as the disk proposal. Following \cite{Golterman:2024ccm}, it was pointed out in \cite{Kaplan:2024ezz}, that the flow of the gauge field is crucial in both proposals of chiral gauge theories. In this paper, our focus is on the slab (Grabowska-Kaplan) proposal \cite{Grabowska:2015qpk}. 

While the proposals for flow were outlined in \cite{Grabowska:2015qpk} using the continuum field theory language, they are yet to be employed on the lattice. The goal of this paper is to explicitly construct the flow for the slab proposal. Furthermore, reference \cite{jansen} in 1992 established how anomaly inflow and anomaly cancellation work for the conventional domain wall fermions on the lattice, where one works with $2n$ dimensional gauge fields that are independent of the extra dimension $x_{2n+1}$. In this paper,  we examine the same in the context of slab geometry with gradient flow \cite{Grabowska:2015qpk} i.e. establish how $2n+1$ current conservation and  anomaly inflow are realized. Interestingly, the equation of motion (EOM) flow which was originally proposed for the disk proposal can also be used to decouple the mirror fermions in the slab proposal \cite{Kaplan:2023pxd}. Thus, we also  explicitly construct the EOM flow for the slab proposal and demonstrate how current conservation ties into anomaly inflow in this case. We address these questions in the context of the disk proposal in a companion paper \cite{PaperB}. Note, that in this paper we will set $n=1$, i.e. construct the flow, establish current conservation and anomaly inflow with a $1+1$ dimensional target chiral gauge theory in mind. The generalization to higher dimensions is straightforward.  

The organization of the paper is as follows. In section \ref{sec:bg} we will use a Euclidean continuum $\mathbb{T}^3$ manifold to review the conventional domain wall fermion
and (i) how it is  combined with extra dimension independent gauge fields to engineer vector-like theories and then (ii) how it is combined with gradient flow in the chiral gauge theory proposal in \cite{Grabowska:2015qpk}.We then formulate a new flow prescription: EOM (equation of motion) flow on the slab, which has not been formulated in the continuum or on the lattice thus far. The EOM flow was proposed in the disk formulation of chiral gauge theories in the continuum \cite{Kaplan:2024ezz, Kaplan:2023pxd} and has been realized on the lattice in the companion paper \cite{PaperB}. The EOM flow constructed for the slab in this paper follows analogous logic.     
In section \ref{sec:lat}, the three types of gauge fields (i) extra dimensions independent, (ii) gradient flow and (iii) EOM flow, are implemented on the lattice. The lattice realization of (i) was performed in \cite{jansen}, and simply helps us understand the main results of the paper concerning the latter two, i.e. gradient flow and EOM flow. 
In section \ref{sec:inflow} we calculate the current density for a specific boundary gauge field which then helps us illustrate the physics of anomaly inflow and anomaly cancellation on the lattice. 

\section{The setup}
\label{sec:bg}
We work in Euclidean space-time dimensions. In this section we will introduce some of the core ideas on which this work is based. For this illustrative purpose, continuous space-time suffices and we will only introduce lattice in the following section. \\

\noindent
{\bf Massless Dirac fermion coupled to $U(1)$ gauge field:}
Here, we first illustrate the simplest vector gauge theory construction using domain wall fermions. The target theory of interest has a single massless Dirac fermion of charge $1$ coupled to $U(1)$ gauge field in $1+1$ dimensions and the associated Lagrangian is given by
\beq
\mathcal{L}_v=\bar{q}\gamma_{\mu}D_\mu q + \frac{1}{4}F_{\mu\nu}F_{\mu\nu}. 
\label{vec}
\eeq
Here $q$ is the massless Dirac fermion in $1+1$ dimensions, $A_\mu$ stands for the $U(1)$ gauge field and $\mu, \nu=t, x$. $\gamma_\mu$ are $1+1$ dimensional Euclidean gamma matrices and  
$D_\mu=\partial_\mu+ i A_\mu$ the covariant derivative. Since the Dirac fermion theory is massless, the target theory has global chiral symmetry. We pick a specific basis for the gamma matrices for convenience, i.e. $\gamma_t=\sigma_1$ and $\gamma_x=\sigma_2$ where $\sigma_{1/2/3}$ are Pauli matrices gives by
\beq
\sigma_1=\begin{pmatrix}
0 && 1\\
1 && 0
\end{pmatrix},\,\, \sigma_2=\begin{pmatrix}
0 && -i\\
i && 0
\end{pmatrix},\,\, \sigma_3=\begin{pmatrix}
1 && 0\\
0 && -1
\end{pmatrix}.
\eeq
We split the Dirac field in left and right chirality components $q=\{q_R, q_L\}$ to rewrite this action as

\beq
\mathcal{L}_v&=&\bar{q}_R (D_t+i D_x) q_R + \bar{q}_L (D_t-i D_x) q_L \nonumber\\
&&+ \frac{1}{4}F_{\mu\nu}F_{\mu\nu}. 
\label{vec2}
\eeq
It is well known that a naive lattice discretization of this Lagrangian leads to fermion doubling \cite{Nielsen:1980rz, Nielsen:1981hk, Nielsen:1981xu, Karsten:1981gd} and does not possess the right continuum limit. However, it was shown in \cite{Kaplan:1992bt} that the doubling problem can be avoided using the extra dimension approach or domain wall fermions. In other words,
 the massless $1+1$ dimensional lattice Dirac fermion theory is realizable as a low energy EFT of a $2+1$ dimensional theory through a domain wall construction \cite{Kaplan:1992bt}. The three directions are denoted as $(t,x, s)$, $s$ being the extra dimension (we substitute $x_1\rightarrow t, x_2\rightarrow x, x_3\rightarrow s$ in the $2+1$ dimensional manifold). We consider a finite manifold in $s$ of length $L_s$ with periodic boundary condition. 
 To engineer the low energy EFT for the fermion sector of Eq. \ref{vec2}, one begins with a $2+1$ dimensional massive Dirac fermion, where the mass of the fermion $m$ is taken to be $s$ - dependent with domain wall defect given by 
\begin{equation}
    m(s) \equiv m_0\Theta(s)
    \label{massf}
\end{equation}
with $m_0>0$ and $\Theta(s)$ is a unit step function given by 
\beq
\Theta(s)= \begin{cases}
-1 \,\,\,\,\,\,\,\text{for}\,\,\,\,\, \frac{L_s}{2}\geq s\geq 0\nonumber\\
1\,\,\,\,\,\,\,\text{for}\,\,\,\,\, -\frac{L_s}{2}<s<0.\nonumber\\
\end{cases}
\eeq
 This mass defect gives rise to a low energy EFT, containing a $1+1$ dimensional left chirality Weyl fermion localized at $s=0$ and a right chirality Weyl fermion localized on $s=L_s/2$. To obtain the target theory including gauge fields, one couples the $2+1$ dimensional Dirac fermion theory to $2+1$ dimensional gauge fields $\bar{A}_i$ where $i=t,s,x$ with the constraint
$\bar{A}_s=0$ and $\bar{A}_{t/x}$ are $s$ independent. We define $\bar{A}_{t/x}(s=0)\equiv A_{t/x}$. The $s$-independence of the gauge fields is necessitated by the vector-like nature of the two dimensional target theory. More generally though, when designing a two dimensional gauge theory coupled to matter within a $2+1$ dimensional construction like here, the requirement is to merely disallow independent gauge field fluctuation for $\bar{A}_{t/x}(s_1)$ and $\bar{A}_{t/x}(s_2)$ for $s_1\neq s_2$. This is because allowing such fluctuations will result in multiple copies of two dimensional fluctuating gauge fields which is undesirable. Thus, the construction of a single fluctuating two dimensional target gauge field only requires that 
 $\bar{A}_{t/x}(s)$ be some function of $\bar{A}_{t/x}(s=0)$ (or $A_{t/x}$). The $s$-independent gauge field for a vector-like theory is a special case of this.

In addition, for the gauge field fluctuation one adds a $1+1$ dimensional gauge field kinetic term, resulting in the full microscopic theory 
\beq
\mathcal{L}_{2+1}&=&\bar{\psi}(\gamma_i D_i+m(s) )\psi\nonumber\\
&+&\frac{1}{4}F_{\mu\nu}F_{\mu\nu}
\label{mic}
\eeq
where $\psi$ is $2+1$ dimensional Dirac fermion with the mass defect of Eq. \ref{massf}, $D_i$ is the $2+1$ dimensional covariant derivative with $i=t,s,x$ and $D_i=\partial_i+i \bar{A}_i$ with constraints on gauge fields as mentioned. 

\begin{figure}
    \centering
\includegraphics[width=.53\textwidth]{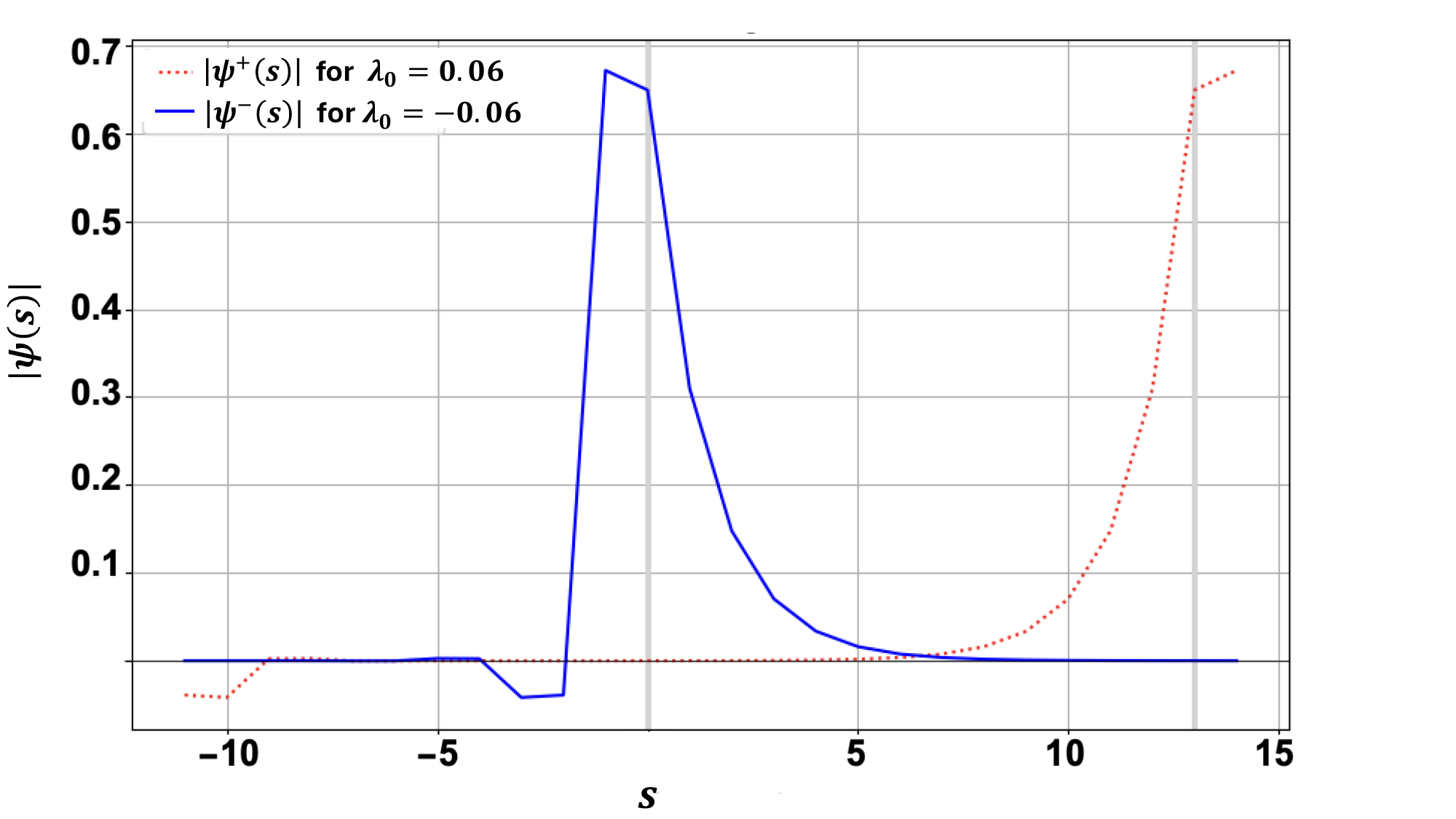}
    \caption{We plot the wavefunction of the lowest lying eigenstates (with eigenvalues $\pm 0.06$) of the Hamiltonian with $ L_x= L_s = 26$, $\mu_0=0.972, w=5.66$. And $\psi^+$ represents a right chirality mode and $\psi^-$ represents left.}
    \label{fig:wavefunction26}
\end{figure}
The right chirality fermion and the left chirality fermion of the target theory in Eq. \ref{vec2} by themselves suffer from the chiral anomaly. Thus, with the definition of chirality projectors
\beq
P_+&=&\frac{1+\sigma_3}{2}\nonumber\\
P_-&=&\frac{1-\sigma_3}{2}
\eeq
the current densities $j^L_{\mu}\equiv \bar{q}(P_+\gamma_\mu P_-) q$ and $j^R_\mu\equiv \bar{q} (P_- \gamma_\mu P_+) q$
are not conserved in the presence of gauge fields. The same must be true of the left and right chirality fermions localized at $s=0$ and $s=L_s/2$ in the higher dimensional microscopic theory of Eq. \ref{mic}. More specifically, from the chiral anomaly we expect, 
\beq
&&\partial_\mu(j^L_\mu-j^R_\mu)= -\frac{Q E}{\pi},\nonumber\\
&&\partial_\mu(j^L_\mu+j^R_\mu)= 0
\label{dj}
\eeq
where $E$ is the electric field, $E=-\partial_t A_x+\partial_x A_t$. Note that $2+1$ dimensional fermion current, $j^{2+1}_i=\bar\psi \gamma_i\psi$ is exactly conserved and can be related to the two-dimensional currents $j_{\mu}^{L/R}$ through 
\beq
j_\mu^L=\int_{\lambda_{\psi}} ds\,\, j_\mu^{2+1}\nonumber\\
j_\mu^R=\int_{\hat\lambda_{\psi}} ds\,\, j_\mu^{2+1}
\eeq
where $\lambda_\psi$ is the region around $s=0$ supporting the width of the left chirality Weyl mode there and $\hat\lambda_\psi$ is the region around $s=L_s/2$ supporting the width of the right chirality Weyl mode. Additionally, $2+1$ dimensional current conservation implies that 
\beq
\partial_\mu j_\mu^L=-\int_{\lambda_{\psi}} ds\,\, \partial_s j_s^{2+1}\nonumber\\
\partial_\mu j_\mu^R=-\int_{\hat\lambda_{\psi}} ds\,\, \partial_s j_s^{2+1}
\eeq

Away from $s=0, L_s/2$, one can integrate out the bulk fermions while including a Pauli Villars regulator. 
With a PV mass $M_{\text{PV}}<0$, this leads to a current \cite{Callan:1984sa}
\beq
j^{2+1}_i=\frac{Q}{4\pi}\epsilon_{ijk}\bar{F}_{jk}\frac{\left(1+\Theta(s)\right)}{2}
\label{cs}
\eeq
when the cutoff (the domain wall height and the PV mass) is taken to infinity. Note that, in our convention $\epsilon_{txs}=1$. 
Here $\bar{F}_{jk}\equiv \partial_j \bar{A}_k-\partial_k \bar{A}_j$, i.e. field strength evaluated using $\bar{A}$ fields. 
The current in Eq. \ref{cs} is referred to as the Chern-Simons current or the Goldstone-Wilczek current for the bulk theory. There is an analogous calculation on the lattice which arrives at the same result \cite{Golterman:1992ub}. Note that, the sharp step function in Eq. \ref{cs} relies on the cutoff being taken to infinity at which point the chiral mode on the boundary has a delta function support. 
When the cutoff is finite, the support for the boundary chiral mode spreads and the expression in Eq. \ref{cs} will be valid outside of the region where the mode has support.

Then it follows that 
\beq
\int_{\lambda_{\psi}} ds\,\, \partial_s j_s^{2+1}&=&\frac{Q \bar{E}(s=0)}{2\pi}\nonumber\\
-\int_{\hat\lambda_{\psi}} ds\,\, \partial_s j_s^{2+1}&=&\frac{Q \bar{E}(s=\frac{L_s}{2})}{2\pi}
\eeq
For $s$-independent gauge fields, $\bar{E}(s=0)=\bar{E}(s=\frac{L_s}{2})=E$ leading to the two equations in Eq. \ref{dj}. Note that, for $s$-independent gauge fields, and with $A_s=0$, the $2+1$ dimensional magnetic field is identically zero. In such a scenario, the only nonzero component of the CS current Eq. \ref{cs} is $j_s^{2+1}$.

\noindent
{\bf The $3-4-5-0$ model:} 
In this paper, we are ultimately interested in chiral gauge theories. The prototypical example in $1+1$ dimension is the $3-4-5-0$ model \cite{Zeng:2022grc}. This will be the focus of our analysis in the rest of this paper. However, most of the concepts illustrated in this paper will hold for any chiral gauge theory (CGT) in any number of dimensions in general. 
The target theory model includes two left chirality Weyl fermion of charges $3$, $4$ and two right chirality ones of charge $5$ and $0$ interacting with a $U(1)$ gauge field \footnote{reversing the chiralitites of the all the fermion species will also produce a chiral gauge theory.}
. The target theory action has the form: 
\beq
\mathcal{L}_c&=&\sum_{Q=5,0}\bar{q}_{R,Q} (D_t+i D_x) q_{R,Q}\nonumber\\
&+& \sum_{Q=3,4}\bar{q}_{L,Q} (D_t-i D_x) q_{L,Q} \nonumber\\
&+& \frac{1}{4}F_{\mu\nu}F_{\mu\nu}
\label{cgt}
\eeq
where the subscript $Q$ on the fermion fields stand for their charge. 

To engineer the fermion content of the theory using the extra dimensional domain wall setup, one can begin with four species of massive Dirac fermions $\psi_Q$ in $2+1$ Euclidean dimensions as before. The subscript $Q$ stands for the charge of the fermion. The masses of the fermions have s-dependence  
of the form 
\beq
m_{3/4}(s)=m_0\Theta(s)
\label{def1}
\eeq
where $m_{3/4}$ are the masses for charge $Q=3, 4$ 
and 
\beq
m_{5/0}(s)=-m_0\Theta(s)
\label{def2}
\eeq
$m_{5/0}$ are the masses for charge $Q=5, 0$. This produces two left chirality Weyl fermions of charge $3$ and $4$ and two right chirality Weyl fermions of charge $5$ and $0$ localized on the wall at $s=0$. 
While this is all we need for the chiral gauge theory, we also end up with a mirror sector of two right chirality fermions of charges $3, 4$ and two left chirality fermions of charges $5$ and $0$ at the anti-wall at $s=L_s/2$. 
Thus, the domain wall construction has twice the fermion content as that of the desired chiral gauge theory.
We eventually want to couple the domain wall setup to gauge fields so as to engineer the target theory of Eq. \ref{cgt}. A natural extension of Eq. \ref{mic} would lead to  \beq
\mathcal{L}_{2+1}=\sum_{Q=3,4,5,0}\bar{\psi}_{Q}(\gamma_i D_i+m_Q(s) )\psi_Q+\frac{1}{4}F_{\mu\nu}F_{\mu\nu}
\label{mic2}
\eeq
where we use $D_i=\partial_i+i Q \bar{A}_i$
with appropriate constraints on $\bar{A}_i$ which we discuss next.

Note that, we have $2+1$ dimensional current densities for each species which we can denote as $j_{i}^{Q,2+1}$. From these one can construct two dimensional currents localized on the wall and the anti-walls, denoted as
$j_{\mu}^{(w), Q, L/R}$ and $j_{\mu}^{(aw), Q, L/R}$ through 
\beq
j_\mu^{(w),Q,L}=\int_{\lambda_{\psi}} ds\,\, j_\mu^{Q, 2+1}\,\,\,\,\text{for $Q=3,4$}\nonumber\\
j_\mu^{(w),Q,R}=\int_{\lambda_{\psi}} ds\,\, j_\mu^{Q, 2+1}\,\,\,\,\text{for $Q=5,0$}\nonumber\\
\label{wc}
\eeq

and 
\beq
j_\mu^{(aw),Q,L}=\int_{\hat\lambda_{\psi}} ds\,\, j_\mu^{Q, 2+1}\,\,\,\,\text{for $Q=5,0$}\nonumber\\
j_\mu^{(aw),Q,R}=\int_{\hat\lambda_{\psi}} ds\,\, j_\mu^{Q, 2+1}\,\,\,\,\text{for $Q=3,4$}.\nonumber\\
\label{awc}
\eeq
The superscripts on the current are self evident, $(w), (aw)$ standing for wall and anti-wall, $Q$ denoting the charge and $L, R$ denoting the chirality.
Chiral anomaly then dictates that
\beq
\partial_\mu j_\mu^{(w),Q,L}&=&-\frac{Q\bar{E}(s=0)}{2\pi}\,\,\,\, \text{for}\,\,\,\, Q=3,4\nonumber\\
\partial_\mu j_\mu^{(w),Q,R}&=&\frac{Q \bar{E}(s=0)}{2\pi}\,\,\,\, \text{for}\,\,\,\, Q=5,0
\label{w2}\eeq
and
\beq
\partial_\mu j_\mu^{(aw),Q,L}=-\frac{Q \bar{E}(s=\frac{L_s}{2})}{2\pi}\,\,\,\, \text{for}\,\,\,\, Q=5,0
\nonumber\\
\partial_\mu j_\mu^{(aw),Q,R}=\frac{Q \bar{E}(s=\frac{L_s}{2})}{2\pi}\,\,\,\, \text{for}\,\,\,\, Q=3,4.\label{aw2}\eeq
Thus, we find that 
\beq
\sum_{Q=3,4}Q(\partial_\mu j_\mu^{(w), Q, L})+\sum_{Q=5,0}Q(\partial_\mu j_\mu^{(w), Q, R})=0
\label{wall}
\eeq
and 
\beq
\sum_{Q=5,0}Q(\partial_\mu j_\mu^{(aw), Q, L})+\sum_{Q=3,4}Q(\partial_\mu j_\mu^{(aw), Q, R})=0
\label{awall}
\eeq
which is just anomaly cancellation in play.
\\ 
\begin{figure*}
    \centering
\includegraphics[width=.8\textwidth]{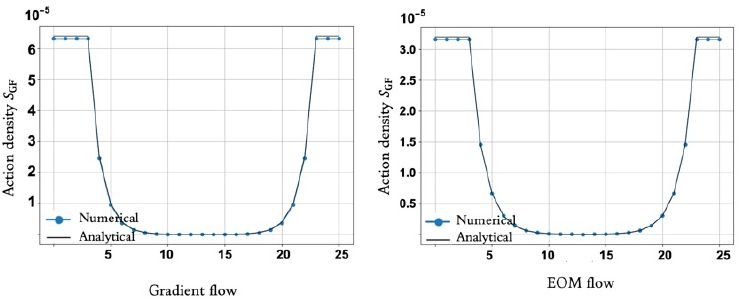}
    \caption{
    On the left panel we plot the gauge action density distribution $S_{\text{GF}}$ for gradient flow, for $ \xi=3$ as a function of the extra dimension $s$ for a lattice with $ L_x = L_t = 16, L_s = 26,  E_0 = 0.001$. Black curves represent $S_{\text{GF}}$ computed using the gauge field in Eq. \ref{gf}. On the right panel, we plot the same quantity with EOM flow for the gauge fields. The black curve represents $S_{\text{GF}}$ computed using the analytical solution to the EOM flow from Eq. \ref{eom}.}
    \label{fig:act-den}
\end{figure*}

\noindent
{\bf $s$-independent gauge fields:}
If we choose $s$ independent gauge fields while setting $A_s=0$, we end up with a vector-like theory, i.e. two copies of a chiral gauge theory coupled to gauge fields in such a way that the resulting theory is a vector-like theory. The corresponding target action is 
\begin{widetext}
\beq
\mathcal{L}_c
&=&\frac{1}{4}F_{\mu\nu}F_{\mu\nu}+\sum_{Q=5,0}\bar{q}_{R,Q} (D_t+i D_x) q_{R,Q}+ \sum_{Q=3,4}\bar{q}_{L,Q} (D_t+i D_x) q_{L,Q} +\sum_{Q=5,0}\bar{q}_{L,Q} (D_t-i D_x) q_{L,Q}\nonumber\\
&+& \sum_{Q=3,4}\bar{q}_{R,Q} (D_t-i D_x) q_{R,Q} \nonumber\\. 
\label{cgt2v}
\eeq
\end{widetext}
The anomaly equations are given by Eq. \ref{w2} and \ref{aw2} with $\bar{E}(s=0)=\bar{E}(s=\frac{L_s}{2})=E$. Of course, Eq. \ref{wall} and Eq. \ref{awall} hold irrespective of the gauge field prescription in the extra dimension.

{\bf slab with gradient flow:} Since $s$ independent gauge fields clearly give rise to a vector gauge theory as seen above, a chiral gauge theory construction would necessarily require a different prescription for the gauge fields. Such a prescription was provided in \cite{Grabowska:2015qpk} where the gauge fields follow gradient flow given by
\beq
\partial_s \bar{A}_{\mu}=\frac{\xi \epsilon(s)}{|\Lambda|}\partial_{\lambda}\bar{F}_{\lambda\mu}.
\label{grad1}
\eeq
Here, $\bar{A}_{\mu}$ is subjected to the boundary condition $\bar{A}_{\mu}(s=0)=A_{\mu}$. $\xi$ is a positive number, and $|\Lambda|$ is set to the cutoff scale. On the lattice, it is typically set by the inverse lattice spacing which is often taken to be $1$. The consequence of such a flow is that the physical gauge field degrees of freedom of the two dimensional gauge field configuration at \cite{Grabowska:2015qpk} $s=0$ get exponentially damped as they reach $s=L_s/2$. This implies that the mirror fermions of the wall at $s=L_s/2$ decouple from the physical gauge fields. \footnote{However, in the presence of $n$ instantons and $\bar{n}$ anti-instantons with $n\neq \bar{n}$, the anti-wall does exhibit a net $n-\bar{n}$ instantons for $n>\bar{n}$ or net $\bar{n}-n$ anti-instanton when $n<\bar{n}$. This has interesting implications for the physics of the particle $\eta'$  when the target theory is the full standard model. For reference see \cite{Golterman:2024ccm, Kaplan:2024ezz}.} This leads to the emergence of the $3-4-5-0$ chiral gauge theory as in Eq. \ref{cgt} as the low energy EFT of the $2+1$ dimensional slab construction.

Interestingly, with any nonzero two dimensional electric field $E\neq 0$ at $s=0$, the field strength at $s\neq 0$ will be $s$ dependent and in fact will be exponentially suppressed with increasing $|s|$. If we assume a strong enough flow, the two dimensional electric field at $s=L_s/2$, $\bar{E}\left(\frac{L_s}{2}\right)$ is approximately zero due to the exponential suppression over $L_s/2$. As a result, the anomaly equations \ref{w2} and \ref{aw2} hold with the substitution
\beq
\bar{E}(s=0)\rightarrow E\nonumber\\
\bar{E}(s=\frac{L_s}{2})\rightarrow 0.
\eeq

There is an additional interesting feature of $s$-dependent gauge fields. In this case, the $2+1$ dimensional magnetic field is nonzero in the bulk. As a result, the expression of Eq. \ref{cs} can  produce nonzero values for both $j_s^{2+1}$ and $j_t^{2+1}$  in the bulk. We will confirm these insights on the lattice in the next section.

{\bf Equation of motion (EOM) flow:} While, gradient flow was suggested specifically in \cite{Grabowska:2015qpk}, a slightly different flow, known as equation of motion flow (EOM flow) was proposed in \cite{Kaplan:2023pvd, Kaplan:2024ezz} in the context of the disk construction. It can be also employed to the slab and we do so here. 
Here the gauge fields in the extra dimension simply follow a higher dimensional equation of motion, which in this case would be $2+1$ dimensional Maxwell's equations subjected to the relevant boundary conditions with $\bar{A}_s=0$. In addition, the solution should also be a stable local minimum of the corresponding $2+1$ dimensional Maxwell action \cite{PaperB}. The impact of this flow on the mirror fermions is similar to that of the gradient flow. The EOM flow exponentially damps out the physical gauge field configurations as a function of $|s|$, just as in the case of gradient flow, thus decoupling the mirror fermions from the gauge field. 

For the purpose of prescribing the gauge field $\bar{A}$ in the interior or the bulk, i.e. $s\neq 0$, we have already restricted ourselves to $\bar{A}_s=0$ sector of the Maxwell action. It may be interesting to consider why this choice maintains the necessary gauge invariance for the problem of interest. 
 The full three dimensional  Maxwell action with unrestricted $\bar{A}_s$ of course respects $2+1$ dimensional gauge invariance. With setting $\bar{A}_s=0$, the gauge freedom reduces to $1+1$ dimensional. However, this is sufficient for our application.

Thus, we propose the modified Maxwell action 
\beq
S_{\text{MW}}^{\text{mod}}=\int d^3 x\,\, \left(\frac{1}{4}\bar{F}_{\mu\nu}\bar{F}_{\mu\nu}+\frac{1}{2}(\partial_s \bar{A}_\mu)^2\right) 
\label{Mwmod}
\eeq
which preserves $1+1$ dimensional gauge invariance in $x$ and $t$. We can now use Euler Lagrange equations corresponding to this action to get to the flow equation:
\beq
\partial_s^2 \bar{A}_\mu + \partial_\nu(\partial_\nu \bar{A}_\mu - \partial_\mu \bar{A}_\nu)=0. 
\label{eomf}
\eeq
This is the equation we use in EOM flow while restricting ourselves to a stable local minimum of the action in Eq. \ref{Mwmod} to extend the gauge field on the wall into the bulk (i.e. for $s\neq 0$).

\section{Lattice construction}
\label{sec:lat}
We will now begin with our lattice analysis for a lattice of size $L_t\times L_x \times L_s$ in $t,s,x$ directions. We set the lattice spacing to one.
The lattice version of the fermion action in Eq. \ref{mic2}  is given by
\begin{equation}
    \begin{aligned}
        S &= \frac{1}{2}\sum_{n,i, Q} \left( \bar{\psi}^n_{Q} \gamma_\mu \left[U_{i}(n)\right]^Q \psi^{n+\hat{i}}_Q - 
    \bar{\psi}^{n+\hat{i}}_Q \gamma_\mu \left[U^*_{i}(n)\right]^Q \psi^{n}_Q \right)\\
    &+ m_Q(s)\sum_n \bar{\psi}^n_{Q} \psi^n_{Q}+ \\
    & w \sum_{n,i} \left( -2\bar{\psi}^n_Q \psi^n_Q + \bar{\psi}^n_Q [U_{i}(n)]^Q \psi^{n+\hat{i}}_Q +
    \bar{\psi}^{n+\hat{i}}_Q [U^*_{i}(n)]^Q \psi^n_Q \right) \\
    & 
    \end{aligned}
    \label{eq:action}
\end{equation}
where the subscript $Q$ stands for the four species with charge $Q$, $i \in \{ t, s, x \}$, $n$ stands for lattice sites taking values in $\{t,s,x\}$
\begin{equation}
    n \in \Lambda = \{(t,s,x)| t \in  \mathbb{Z}_{L_t}, x \in \mathbb{Z}_{L_x},s \in \mathbb{Z}_{L_s} \}
\end{equation}
and $w$ is the Wilson parameter.
$U_i(n)$ is the Wilson line in the direction $i$ on gauge link associated with lattice site $n$. 
We exclude the gauge action since the the goal of this paper is to understand current conservation in a background gauge field. 
For the fermions, we use periodic boundary conditions (PBC) in $s$ and $t$ and anti-periodic boundary condition (APBC) in $x$.
For $Q=3,4$, the mass defect on the lattice is given by
\begin{equation}
    m_Q(s) \equiv \begin{cases}
        -\sinh(\mu_0) & 1 \leq s < L_s/2 \\
        +\sinh(\mu_0) & L_s/2 +1 \leq s < L_s \\
        0, & s = 0, L_s/2 ,
    \end{cases}
    \label{mass}
\end{equation}
with $\mu_0>0$.
For $Q=5,0$ the sign of the mass term is flipped. Our goal is to understand current conservation in background gauge fields for which we need to consider fermion coupling to the gauge fields. Before we do this exercise, we turn the gauge fields off, set the parameters $\mu_0 = 0.972$ and $w = 0.566$, and solve the Hamiltonian corresponding to Eq. \ref{eq:action} for a lattice size of $L_x=16, L_t=16, L_s=26$. We find the two lowest eigenvalues and the corresponding wave functions. We plot the magnitude of the wave function in Fig. \ref{fig:wavefunction26} which shows that the two wave-functions are localized on the wall and the anti-wall spanning approximately $7$ sites in $s$ around the two defects. 
\begin{figure*}
    \centering
\includegraphics[width=1.01\textwidth]{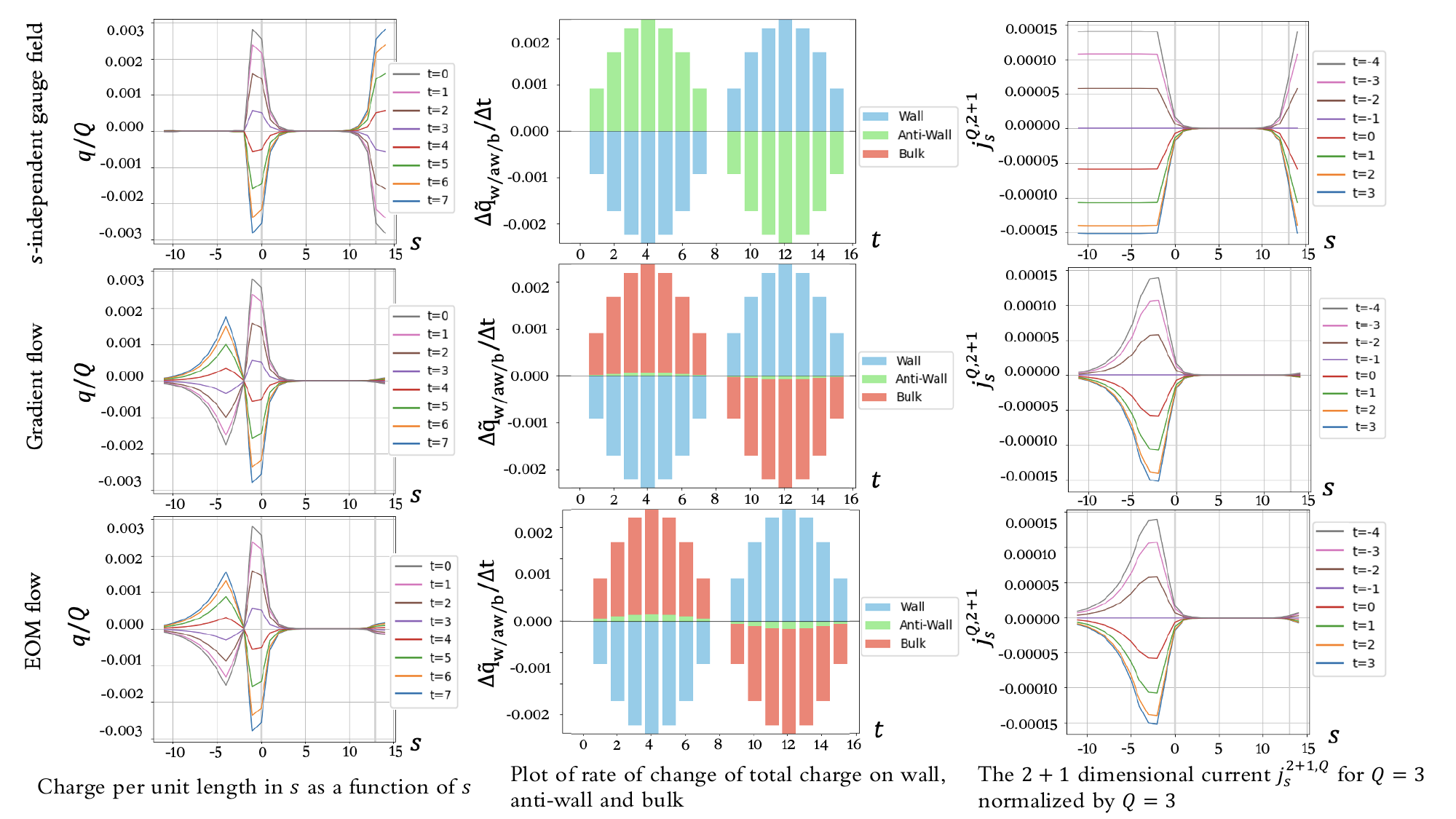}
    \caption{We plot, charge per unit length as defined in Eq. \ref{pul}, the rate of change of the total charge in wall, anti-wall and bulk and $s^{\text{th}}$ component of the $2+1$ dimensional current $j_{s}^{2+1, Q}$ normalized to $Q$ for $Q=3$ fermion with $s$-independent gauge field for
    $L_x = L_t = 16$, $L_s = 26$ , $w = 0.566$, $m_0 = 1.132$,
    $\mu_0 = 0.972$, and $E_0 = 0.001$. }
    \label{fig1}
\end{figure*}

\begin{figure*}
    \centering
\includegraphics[width=1.1\textwidth]{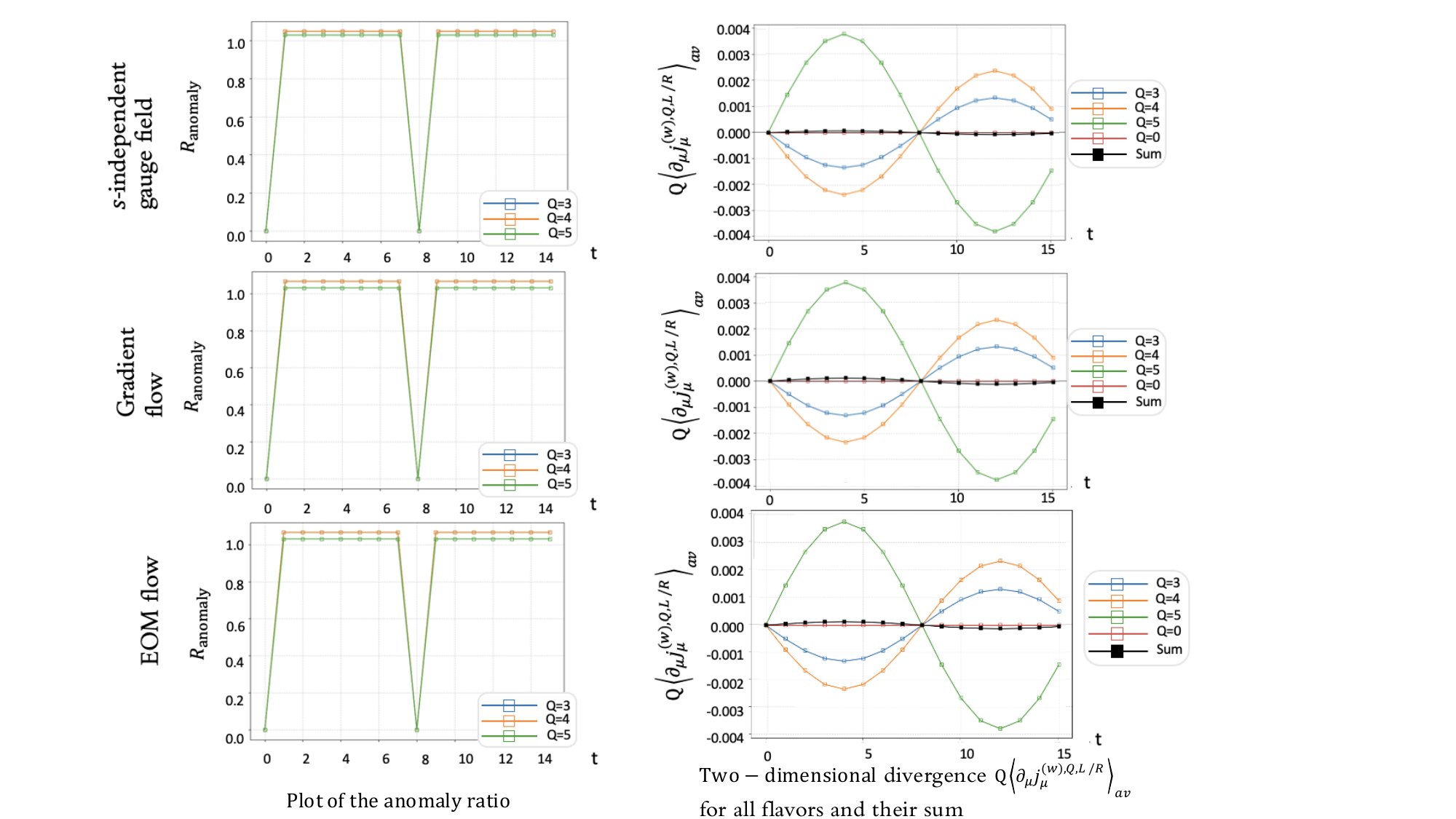}
    \caption{We plot the anomaly 
    ratio $R_{\mathrm{Anomaly}}$ which is $\approx 1.045$ for $Q=3, 4$ and $\approx 1.029$ for $Q=5$. We also plot $Q \langle \partial_\mu j_\mu^{(w),Q,L/R}\rangle_{\text{av}}$ for the four species with charges $Q=3,4,5,0$ as well as their sum to show anomaly cancellation.
    }
    \label{fig2}
\end{figure*}

In the following discussion, we will specialize to a specific boundary gauge field configuration for $A_{x,t}$. This configuration was first used by \cite{jansen} to illustrate current conservation and anomaly inflow for conventional domain wall fermions. We use the same configuration here as a representative gauge field configuration. The analysis presented here extends straight forwardly to any other gauge field configuration.  

\noindent
{\bf A comment on fermion operator:} Note that one can extract the fermion operator for the charge $Q$ fermion from Eq 28. We can denote it as $D^Q$ and write the action as
\beq
S=\sum_{n, m, Q}\bar{\psi}^n_Q D_{nm}^Q \psi^m_Q.
\eeq
For the purpose of this work it is perfectly reasonable to use $D^Q$ as the fermion operator. However, for a chiral gauge theory application involving dynamical fermions and gauge fields \cite{Kaplan:2024ezz} suggested replacing this fermion operator by a modified fermion operator $\tilde{D}^Q$ given by
\beq
\tilde{D}^Q=\frac{D^Q}{\sqrt{D^Q(D^Q)^{\dagger}+\mu^2 \delta(s)+\epsilon}}
\eeq
where $\mu$ is some mass scale and $\delta(s)$ has support only at $s=0$ and $\epsilon$ is a small scale with a dimension of mass squared to be taken to zero at the end of the calculation. 
The long wavelength chiral modes of the two operators are analogous and thus their anomaly inflows and anomaly cancellation physics should be analogous. Thus, we refrain from using $\tilde{D}^Q$ for this analysis. However, for a simulation with dynamical fields, $\tilde{D}^Q$ is required to remove the real part of the radiative corrections to the gauge field action originating from the bulk fermions \cite{Kaplan:2024ezz}.

\subsection{$s$-independent gauge fields}
As mentioned before $s$-independent gauge field configurations are used in conventional domain wall setup to simulate vector-like theories with good global chiral symmetry. Following \cite{jansen} we take the boundary field configuration to be 
\beq
    A_x&=&  \frac{L_t}{2\pi} E_0 \cos\left(\frac{2\pi}{L_t} (t )\right),\nonumber\\\nonumber\\
    A_t&=& 0.\nonumber\\
    \label{eq:nff}
\eeq
This leads to the following profile for the higher dimensional or $\bar{A}$ gauge fields
\beq
    \bar{A}_x&=&  \frac{L_t}{2\pi} E_0 \cos\left(\frac{2\pi}{L_t} (t )\right),\nonumber\\\nonumber\\
    \bar{A}_t&=& 0\nonumber\\
    \bar{A}_s&=& 0.
    \label{eq:nf}
\eeq
Since this ansatz was analyzed in \cite{jansen}, we will replicate those results here. Our goal is to contrast the results of gradient flow and EOM flow with that of $s$-independent gauge field.

For this gauge field, using Eq. \ref{cs} we note that for every species, we expect a CS current of    
\beq
j_s^{Q, 2+1}&=&\frac{Q}{2\pi}\bar{E}\frac{(1+\Theta(s))}{2}\nonumber\\
&\approx &-\frac{Q E_0}{2\pi} \sin\left(\frac{2\pi}{L_t} (t )\right)\frac{(1+\Theta(s))}{2}\nonumber\\
j^{Q,2+1}_t&=&j^{Q,2+1}_x=0.
\eeq
We use the approximate sign above since in this equation we have used continuous derivatives instead of finite difference to relate gauge field to electric field. In the continuum limit, the two natch upto small discretization errors.

\subsection{Gradient flow}
As suggested in \cite{Grabowska:2015qpk} the gradient flow equation is given by Eq. \ref{grad1}.

Our goal in this paper is to focus on aspects of current conservation and compare and contrast the results of gradient flow and EOM flow with the $s$-independent gauge field. To, make the comparison concrete with the gauge field in Eq. \ref{eq:nf} we will set 
\beq
    A_x(t)&=&  \frac{L_t}{2\pi} E_0 \cos\left(\frac{2\pi}{L_t} t\right),\nonumber\\\nonumber\\
    A_t(s=0)&=& 0. 
    \label{eq:bc}
\eeq
For the $\bar{A}$ gauge fields, we already know that 
\beq
\bar{A}_s=0 
\label{as}
\eeq
as set by all the gauge field proposals we have discussed so far. 

To get an idea of what the flowed $\bar{A}_{t,x}$ gauge fields will look like, we can solve the analytic gradient flow equation, Eq. \ref{grad1} subject to the boundary condition of Eq. \ref{eq:bc}. However, we modify Eq. \ref{grad1} slightly to take into account the fact that the low lying fermion boundary modes have a finite width in $s$ as seen in Fig. \ref{fig:wavefunction26}. We have been denoting this width as $\lambda_\psi$. Thus, we keep the gauge field $s$-independent within the region $\frac{\lambda_\psi}{2}\geq s\geq -\frac{\lambda_\psi}{2}$ and flow beyond this.

This leads to gauge field solutions:
\beq
    \bar{A}_x &= & 
\frac{L_t}{2\pi} E_0 \cos\left(\frac{2\pi}{L_t} (t )\right) \nonumber\\
&&\left(1-H(s)+e^{-\xi \left(\frac{2\pi}{L_t}\right)^2 (|s|-\frac{\lambda_\psi}{2}) }H(s)\right)\nonumber
\\
    \bar{A}_t &=& 0\nonumber\\
\bar{A}_s&=& 0\nonumber\\
\label{gf}
\eeq
where 
\beq
H(s)=\begin{cases} 0 \,\,\,\,\text{for} \frac{\lambda_\psi}{2}\geq s\geq \frac{-\lambda_\psi}{2}\nonumber\\
1 \,\,\,\, \text{otherwise.}
\end{cases}\nonumber\\
\label{h}
\eeq
\begin{widetext}
We can compute the corresponding $2+1$ dimensional field strengths using continuous derivatives as 
\beq
&&\bar{F}_{tx}\approx  -E_0 \sin\left(\frac{2\pi}{L_t} (t )\right) \left(1-H(s)+ e^{-\xi \left(\frac{2\pi}{L_t}\right)^2 (|s|-\frac{\lambda_\psi}{2})}H(s)\right)=-\bar{E}, \nonumber\\
        &&\bar{F}_{sx} \approx\begin{cases} \left(- H(s)\xi \left( \frac{2\pi}{L_t} \right)^2e^{-\xi \left(\frac{2\pi}{L_t}\right)^2 (|s|-\frac{\lambda_\psi}{2})  }\right) E_0 \cos\left(\frac{2\pi}{L_t} (t )\right)  \,\,\,\,\text{for $s>0$}\nonumber\\
        \left( H(s)\xi \left( \frac{2\pi}{L_t} \right)^2e^{-\xi \left(\frac{2\pi}{L_t}\right)^2 (|s|-\frac{\lambda_\psi}{2})  }\right) E_0 \cos\left(\frac{2\pi}{L_t} (t )\right)  \,\,\,\,\text{for $s<0$}\end{cases}\nonumber\\\label{eq:flowF}
    \eeq
    \end{widetext}
where $\bar{F}_{tx}$ gives the electric field ($-\bar{E}$) as a function of $t,x, s$ and the magnetic field $\bar{F}_{sx}$ as a function of $t,x, s$. Note that, we use approximate signs in Eq. \ref{eq:flowF} to indicate that to evaluate $\bar{F}$ we have replaced lattice finite difference with continuous derivatives. The results for $\bar{F}_{tx}, \bar{F}_{sx}$ suggest interesting behavior for the Chern-Simons current which in this case will be given by 
\beq
j_s^{Q, 2+1}=\left(\frac{1+\Theta(s)}{2}\right)\frac{Q}{2\pi} \bar{F}_{tx}\nonumber\\
j_t^{Q, 2+1}=\left(\frac{1+\Theta(s)}{2}\right)\frac{Q}{2\pi}\bar{F}_{xs}
\label{eqF}
\eeq
where one would substitute the RHS of Eq. \ref{eq:flowF}. Thus, we see that, $s$-th component of the CS current, i.e. $j_s^{2+1}$ experiences exponential decay beyond $|s|>\frac{\lambda_\psi}{2}$ in contrast with the previous case of $s$-indepdent gauge fields.
Similarly, the CS charge density, i.e. $j_t^{Q, 2+1}$ becomes nonzero at $|s|\approx\frac{\lambda_\psi}{2}$, again in contrast with the $s$-independent case where $j_t^{2+1}$ was identically zero away from $s=0, L_s/2$. Strictly speaking, the expression for the CS current obtained from integrating out the bulk fermions in Eq. \ref{cs} hold sufficiently away from the domain wall and the current in Eq. \ref{eqF} will not produce exact quantitative results for $j^{Q,2+1}$ near $|s|\approx\frac{\lambda_\psi}{2}$. However, they produce the correct qualitative feature as we will see from the numerical analysis later in this paper.

As we extend this to the lattice, note that the region $\lambda_\psi$ approximately extends over the lattice sites $s=-3,-2,-1,0,1,2,3$ for our choice of parameters for $\mu_0$ and $w$. Thus, we will implement gradient flow outside this region while holding the gauge field $s$-independent within this region. 
Eq. \ref{grad1} is the continuum version of the gradient flow. Clearly, for a lattice construction the lattice version is desirable. On a lattice, for $|s|>3$, we apply the following flow equation to the gauge field with $\Lambda = 1$

\begin{equation}
    \partial_s U_\mu = - \xi \{\partial_{x,\mu} \hat{S} \} U_\mu \label{eq:WF0}
\end{equation}

where $\partial_{x,\mu}$ stands for the natural $U(1)$ valued differential operator with respect to the link variable $U_\mu$ as shown in \cite{luscher2010properties}  and $\hat{S}$ is the action given by 
\begin{equation}
    \hat{S} = \sum_{n}\sum_{\mu,\nu}\mathrm{Re}\ \mathrm{tr}\{ 1 - U_\mathrm{plaquette}(n,\mu,\nu) \} 
    \label{hats}
\end{equation}
with  
\beq
U_\mathrm{plaquette}(n,\mu,\nu)
&=& U_\mu(n)U_\nu(n+\hat\mu)\nonumber\\
&&U_\mu(n+\hat\nu)^{-1}U_\nu(n)^{-1}. 
\eeq

$n$ stands for coordinates of lattice sites in $t,x, s$ and $\mu, \nu $ can be $t$ or $x$. It was shown in \cite{luscher2010trivializing}  
\beq
\partial_{x,\mu} \hat{S} =Z[U]
\eeq

where $Z[U]$ is given by 

\begin{equation}
    \begin{aligned}
        &Z[U] =\\ & \sum_{\nu\neq\mu} \mathcal{P}\left\{ 
        U_\mu(n)U_\nu(n+\hat\mu)U_\mu(n+\hat\nu)^{-1}U_\nu(n)^{-1} \right. \\ 
        & \left. + U_\mu(n)U_\nu(n+\hat\mu-\hat\nu)^{-1}U_\mu(n-\hat\nu)^{-1}U_\nu(n-\hat\nu) \right\} \\
    \end{aligned}
\end{equation}
leading to the equation:
\begin{equation}
    \partial_s U_\mu =- \xi Z[U] U_\mu \label{eq:WF}.
\end{equation}
We set $\xi=3$ for our subsequent analysis.

Eq. (\ref{eq:WF}) is referred to as the `Wilson flow' equation  \cite{luscher2010properties}.  The Euler scheme for numerical calculation of the `Wilson flow' equation is 
\begin{equation}
    U_\mu(n+\epsilon\hat{s}) = e^{-\epsilon \xi Z[U]} U_\mu(n) \label{eq:numWF}
\end{equation}
as discussed in  \cite{luscher2010trivializing}, where $\epsilon$ is a small parameter  set as $1/7$ for $\xi=3$ which satisfies $ \epsilon \xi < 1/2 $ (converging condition).

To see the effect of the flow we
define a two dimensional gauge action density as 
\begin{equation}
    S_{\text{GF}}(s) = \frac{1}{2} \sum_{t,x} \mathrm{Re}\{1 - U_{\text{plaquette}}(t,x)\}.
    \label{eq:sgf}
\end{equation}
Note that $S_{\text{GF}}$ is a function of $s$. We substitute the flow gauge field in this expression and plot $S_{\text{GF}}$ as a function of $s$ direction in Fig. \ref{fig:act-den}. We expect the $S_{\text{GF}}$ do exponentially decay as a function of $|S|$. This behavior is confirmed in Fig. \ref{fig:act-den}.

We will now have to substitute this gauge field, in the fermion action to compute the corresponding current density. 
The earlier discussion about CS currents suggest that, we not only expect a nonzero $j_s^{Q,2+1}$, we also expect a nonzero $2+1$ dimensional charge density, i.e. $j_t^{Q,2+1}$. Clearly, if we follow the Eq. \ref{cs} all the way to $s\sim \frac{\lambda_\psi}{2}$, i.e. the border of its regime of validity, a nonzero $j_t^{Q,2+1}$ is tied to a nonzero $2+1$ dimensional magnetic field, both of which are expected to peak near $s=\pm \frac{\lambda_\psi}{2}\approx \pm 3$. As one increases $|s|$, the gauge fields die down exponentially. As a result, the domain wall at $s=L_s/2$ is not expected to experience any electromagnetic fields and does not participate in current conservation.

\subsection{EOM flow} For the EOM flow we begin with the same boundary condition as in Eq. \ref{eq:bc}. We also impose
$\bar{A}_s=0$ as in Eq. \ref{as}. We again keep in mind that the fermion boundary mode extends from
$-\frac{\lambda_\psi}{2}\geq s\geq \frac{\lambda_\psi}{2}$. Thus we keep the gauge field constant within this region as in the case of gradient flow and flow beyond this region. The corresponding gauge field solutions of Eq. \ref{eomf} are: 
\beq
    \bar{A}_x &=& 
\frac{L_t}{2\pi} E_0 \cos\left(\frac{2\pi}{L_t} (t )\right)\nonumber\\
&&\left(1-H(s)+ e^{- \left(\frac{2\pi}{L_t}\right) \left(|s|-\frac{\lambda_\psi}{2}\right) }H(s)\right)\nonumber
\\
    \bar{A}_t &=& 0\nonumber\\
\bar{A}_s&=& 0\nonumber\\
\label{eom}
\eeq
where $H(s)$ is given by Eq. \ref{h}. Note that, there is also a solution to the EOM that grows exponentially with $s$. We have discarded this solution since we want to be at a stable local minimum of the $2+1$ dimensional modified Maxwell action Eq. \ref{Mwmod}.

\begin{widetext}
We can compute the corresponding $2+1$ dimensional field strengths given by  
\beq
&&\bar{F}_{tx}\approx  -E_0 \sin\left(\frac{2\pi}{L_t} (t )\right) \left(1-H(s)+ e^{- \left(\frac{2\pi}{L_t}\right) (|s|-\frac{\lambda_\psi}{2})}H(s)\right), \nonumber\\
        &&\bar{F}_{sx} \approx\begin{cases} \left(- H(s) \left( \frac{2\pi}{L_t} \right)e^{- \left(\frac{2\pi}{L_t}\right) (|s|-\frac{\lambda_\psi}{2})  }\right) E_0 \cos\left(\frac{2\pi}{L_t} (t )\right)  \,\,\,\,\text{for $s>0$}\nonumber\\
        \left(+ H(s)\left( \frac{2\pi}{L_t} \right)e^{- \left(\frac{2\pi}{L_t}\right) (|s|-\frac{\lambda_\psi}{2})  }\right) E_0 \cos\left(\frac{2\pi}{L_t} (t )\right)  \,\,\,\,\text{for $s<0$}\end{cases}\nonumber\\\label{eq:flowF2}
    \eeq
    \end{widetext}
Thus, we see that, beyond $|s|>\frac{\lambda_\psi}{2}$, EOM flow too produces decay in gauge fields in analogy with that in gradient flow Eq. \ref{eq:flowF}. Eq. \ref{eqF} still holds except with expressions of $\bar{F}$ given by Eq. \ref{eq:flowF2}. As with the case for gradient flow, $j_t^{Q, 2+1}$ peaks around $s=\frac{\lambda_\psi}{2}$ and dropping off away from it. And $j_s^{Q, 2+1}$ is nonzero near $s=0$ and dropping off exponentially beyond $|s|>\frac{\lambda_\psi}{2}$. 

To implement Eq. \ref{eomf} on the lattice, we note that it is equivalent to 
\beq
\partial_i \bar{F}_{i\mu}=0
\label{eqeom}
\eeq
with $\bar{A}_s=0$.
A simple way to implement this is to introduce an additional imaginary time direction $\tau$ and allow the gauge field to depend on $\tau$. One then uses gradient flow in $\tau$ as in 
\beq
\partial_\tau \bar{A}_\mu=\xi\partial_i\bar{F}_{i\mu} 
\label{taueq}
\eeq
where $\xi$ is the corresponding gradient flow parameter.
The logic is simply that, after long enough flow time, $\bar{A}$ becomes $\tau$ independent causing the LHS of Eq. \ref{taueq} to go to zero at which point $\bar{A}$ solves the EOM of Eq. \ref{eqeom}. In addition, the gradient flow in this additional time automatically selects the stable local minimum of the $2+1$ dimensional modified gauge action of Eq. \ref{Mwmod}.

On the lattice we implement Eq. \ref{taueq} using 
\begin{equation}
    \partial_\tau U_\mu(n,\tau) = \xi Z_{2+1}[U] U_\mu(n,\tau)
\end{equation}

where in $Z_{2+1}$ is 

\begin{equation}
    \begin{aligned}
        &Z_{2+1}[U] =\\ &- \sum_{j\neq i} \mathcal{P}\left\{ 
        U_i(n,\tau)U_j(n+\hat i, \tau)U_i(n+\hat j, \tau)^{-1}U_j(n, \tau)^{-1} \right. \\ 
        & \left. + U_i(n, \tau)U_j(n+\hat i-\hat j, \tau)^{-1}U_i(n-\hat j, \tau)^{-1}U_j(n-\hat j, \tau) \right\} \\
    \end{aligned}
\end{equation}
and we have made the $\tau$ dependence explicit in our notation.
Also, $i,j \in \{t,s,x\} $ and $n$ stands for lattice sites in ($t,s,x$). We also impose the constraint $U_{s} = 1$ and for links with $ |s| \leq \frac{\lambda_\psi}{2}$, $U_\mu$ are $s$ independent, i.e. $U_\mu(t, s, x, \tau)=U_\mu(t, x, 0, \tau)=U_\mu(t, x, 0, 0)$. To solve the flow equation in $\tau$, we will need an initial gauge configuration at $\tau=0$ for all of $s$. We take $s$-independent gauge fields to be the initial condition. Also, we set the parameter $\xi$ to $0.1$ here. We obtain the solution to the EOM flow after flowing long enough in $\tau$ (after $1000$ steps). We plot the action density of $S_{\text{GF}}$ defined in Eq. \ref{eq:sgf} as a function of $s$ in Fig. \ref{fig:act-den}. Again, we expect to see exponentiall fall as a function of $|s|$ and this is verified in fig. \ref{fig:act-den}.

\section{current density comparison}
\label{sec:inflow}
We will now verify our expectations regarding current conservation and anomaly inflow. The summary of our expectations is that,
\begin{itemize}
\item For the $s$-independent gauge field, we expect charge non-conservation on the wall and the anti-wall. We also expect the charge non-conservation on the wall and the anti-wall to cancel each other. And we expect a an $s$ independent $j_s^{Q,2+1}$ in the bulk $s\neq 0, L_s/2$ . 
\item For the gradient flow and EOM flow, there will be charge non-conservation on the wall at $s=0$ as well as in the bulk. The charge non-conservation of the wall and the bulk will cancel each other. The anti-wall will not participate in charge conservation. The current $j_s^{Q,2+1}$ will be nonzero near $s\sim 0$, dropping off exponentially in $|s|$ beyond where the flow begins, i.e. $|s|=\frac{\lambda_\psi}{2}$. 
\item We expect anomaly equations to hold in each case. 
\end{itemize}

We begin with the current density operator on the lattice given by 
\beq
   && j^{Q, 2+1}_i = \frac{1}{2} [ \bar{\psi}^n_Q \gamma_i [U_{i}(n)]^Q \psi^{n+\hat{i}}_Q +
    \bar{\psi}^{n+\hat{i}}_Q \gamma_i [U_i^*(n)]^Q \psi^n_Q ] \nonumber\\
    &&
    + w [ \bar{\psi}^n_Q  [U_{i}(n)]^Q \psi^{n+\hat{i}}_Q - 
    \bar{\psi}^{n+\hat{i}}_Q [U^*_{i}(n)]^Q \psi^n _Q]. 
    \label{curr}
\eeq
Thus, we can compute the current using 
\begin{equation}
    \langle j^{Q,2+1}_{i} \rangle = \frac{1}{\hat{Z}} \int [D\bar{\psi} D\psi] e^{-S} j_{i}^{Q, 2+1} = -\mathrm{Tr}[\hat{j}_{i}^{Q,2+1} h^{-1}]
\end{equation}
where $h$ is the combined $2+1$ dimensional fermion operator extracted from the action in Eq. \ref{eq:action} for the four species, with charges $Q=3,4,5,0$ in the presence of the gauge field, $\hat{j}_{i}^{Q,2+1}$ is the current density operator extracted from Eq. \ref{curr} and $\hat{Z}$ is the partition function
\beq
\hat{Z}=\int [D\bar{\psi} D\psi] e^{-S} j_{i}^{Q, 2+1}.
\eeq

Folloing Eq. \ref{wc}, we now want to define a two dimensional current centered on the wall at $s=0$ to capture the physics of chiral anomaly for the boundary fermions. 
We begin by defining a two dimensional divergence of the current by backward difference given by
\beq
    \langle \partial_\mu j_\mu^{Q, 2+1} \rangle &\equiv& \left( \langle j_t^{Q,2+1}(n) \rangle - \langle j^{Q,2+1}_t(n-\hat{t}) \rangle \right)\nonumber\\
    &+& \left( \langle j_x^{Q,2+1}(n) \rangle - \langle j_x^{Q,2+1}(n-\hat{x}) \rangle \right) \label{eq:div}
\eeq
where $\mu=t, x$. Following Eq. \ref{wc} and \ref{awc} we sum over $s$ in the regions $\lambda_\psi$ and $\hat{\lambda}_\psi$ to get a sensible two dimensional divergence of two dimensional current 
\beq
    \langle \partial_\mu j_\mu^{(w),Q, L}  \rangle &=&  \sum_{s\in\Lambda_\psi}\langle \partial_\mu j^{Q,2+1}_\mu \rangle\,\,\text{for}\,\, Q=3, 4\nonumber\\
    \langle \partial_\mu j_\mu^{(w),Q, R}  \rangle &=&  \sum_{s\in\Lambda_\psi}\langle \partial_\mu j^{Q,2+1}_\mu \rangle\,\,\text{for}\,\, Q=5, 0\nonumber\\
    \langle \partial_\mu j_\mu^{(aw),Q, R}  \rangle &=&  \sum_{s\in\hat\Lambda_\psi}\langle \partial_\mu j^{Q,2+1}_\mu \rangle\,\,\text{for}\,\, Q=3, 4\nonumber\\
    \langle \partial_\mu j_\mu^{(aw),Q, L}  \rangle &=&  \sum_{s\in\hat\Lambda_\psi}\langle \partial_\mu j^{Q,2+1}_\mu \rangle\,\,\text{for}\,\, Q=5, 0\nonumber\\
    \label{eq:dj2}
\eeq
Since the gauge fields in Eq. \ref{eq:nff}, \ref{gf}, \ref{eom} happen to be $x$-independent, we
take an average in $x$ to define: 
\beq
    \langle \partial_\mu j_\mu^{(w),Q, L/R}  \rangle_{\text{av}} = \frac{1}{L_x}\sum_x \sum_{s\in\Lambda_\psi}\langle \partial_\mu j^{Q,2+1}_\mu \rangle\nonumber\\
     \langle \partial_\mu j_\mu^{(aw),Q, L/R}  \rangle_{\text{av}} = \frac{1}{L_x}\sum_x \sum_{s\in\hat\Lambda_\psi}\langle \partial_\mu j^{Q,2+1}_\mu \rangle.\label{eq:dj}
\eeq
Eq. \ref{w2}, \ref{aw2} should hold for the divergences defined in Eq. \ref{eq:dj} just as they would hold for the ones defined in Eq. \ref{eq:dj2} as long as the gauge fields are $x$ independent.

To verify that the anomaly for the boundary theory at $s=0$ is working correctly we define the ratio $R_{\mathrm{Anomaly}}$, such that

\beq
    &&R_{\mathrm{Anomaly}}(t)\nonumber\\
    &\equiv& ( QE_{\text{eff}}(t)/2\pi ) / \langle \partial_\mu j_\mu^{(w), Q, L} (t) \rangle_{\text{av}}\,\,\text{for}\,\, Q=3,4\nonumber\\
    &=&(- QE_{\text{eff}}(t)/2\pi ) / \langle \partial_\mu j_\mu^{(w), Q, R} (t) \rangle_{\text{av}}\,\,\text{for}\,\, Q=5,0\nonumber\\
    \label{ran}
\eeq
where 
\beq
E_{\text{eff}}(t)&=& \bar{E}(s=0)\nonumber \\ &=&  E_0 \frac{ \sin(2\pi/L_t) } {2\pi/L_t} \sin\left(\frac{2\pi(t)}{L_t}\right) 
\eeq
defined using central lattice finite difference. Note that this is in contrast with previous estimates of the electric field where we used derivatives to relate electric fields to gauge field neglecting discretization errors. There, our goal was to get qualitative understanding of the Chern-Simons current. Here, we use finite difference to define $E_{\text{eff}}$ since our goal is to quantitatively demonstrate anomaly inflow on the lattice.
We expect the ratio in Eq. \ref{ran} to be close to $1$ due to the chiral anomaly as in Eq. \ref{w2}. We could have defined an analogous ratio for the anti-wall at $s=L_s/2$ and confirm the the chiral anomaly results there. For the purpose of this analysis, we will focus on the $s=0$ wall. 
Furthermore, we will define the following to capture the charge density (charge per unit length in $s$ to be precise)
\beq
    q(t,s) = \sum_x \langle j^{Q,2+1}_t(t,s,x) \rangle .
    \label{pul}
\eeq
We will track the behavior of $q(t,s)$ to confirm its behavior for the $s$-independent gauge field case where we expect it to be zero everywhere in the bulk, and that for the gradient flow and EOM flow, when it is expected to peak at the boundary of $\lambda_\psi$. We will also define total charge on wall, anti-wall and bulk by 
\beq
\tilde{q}_{\text{w}}(t) &=& \sum_x\sum_{s\in\Lambda_{\psi}} \frac{q(t,s)}{Q} ,\nonumber\\ 
\tilde{q}_{\text{aw}}(t) &=& \sum_x\sum_{s\in\hat\lambda_\psi} \frac{q(t,s)}{Q},
   \nonumber\\ \tilde{q}_{\text{b}}(t) &=& \sum_x\sum_{s} \frac{q(t,s)}{Q}-\tilde{q}_\text{w}-\tilde{q}_\text{aw}.
\eeq
\\\\

\noindent
{\bf Numerical results for current for the three cases:}
We now compare the behavior of the charge and current density and for the three different prescriptions for the gauge fields and how anomaly inflow works in each case.

In Fig.\ref{fig1} we plot the variation of charge per unit length, normalized to $Q$, i.e. $q(t,s)/Q$ for $Q=3$ on the left panel. We show how it changes with various time slices. As expected, for the $s$-independent gauge field, there is charge accumulation on the wall as well as the anti-wall with time. For the gradient flow and the EOM flow, the anti-wall does not experience any gauge field and and thus no charge gets accumulated or depleted there. Of course, total charge must be conserved. Thus, we see charge accumulation/depletion slightly away from the wall, i.e. right at the boundary of the region $\lambda_\psi$ , or $s=-\lambda_\psi/2$ where the flow starts. This is expected from the analytic results. 

In the middle panel, we plot the rate of change of total charge in the wall, anti-wall and the bulk: $\Delta \tilde{q}_{\text{w}}/(\Delta t)$, $\Delta \tilde{q}_{\text{aw}}/(\Delta t)$ and $\Delta \tilde{q}_{\text{b}}/(\Delta t)$ for flavor $Q=3$ with $\Delta t=1$. For the $s$-independent gauge field we find that the charge depleted from the wall at $s=0$ is being deposited on the wall at $s=L_s/2$ and vice versa. Just as expected, for the gradient flow and the EOM flow the anti-wall does not participate or participates minimally in charge conservation. Instead the charge deposited on the wall is supplied mostly by the bulk. 

On the right panel, we plot the $s^{\text{th}}$ component of the $2+1$ dimensional current, i.e. $j_s^{Q,2+1}$ normalized by $Q$ for $Q=3$. For the $s$-independent gauge field case, this is the Chern-Simons current and is nonzero in the bulk. For the gradient flow and the EOM flow, the current falls off rapidly in the bulk, which is to say that the $s^{\text{th}}$ component of the Chern-Simons current in the bulk is almost zero. There is nonzero $j_s^{Q,2+1}$ near the boundary, $s=0$, which carries charge from the region around $s\approx \lambda_\psi/2 \approx 3$ to $s=0$.

{\bf Numerical results for anomaly inflow and anomaly cancellation for the three cases:} In Fig \ref{fig2} we plot the anomaly ratio 
$R_{\mathrm{Anomaly}}$ as well as the divergence of the two dimensional current density $Q \langle \partial_\mu j_\mu^{(w),Q,L/R}\rangle_{\text{av}}$ for the three prescriptions for the gauge field. In each case, we find that  $R_{\mathrm{Anomaly}}\approx 1$ which confirms that the anomaly inflow is working the way we expect it to at the wall at $s=0$. The weighted sum of the two dimensional divergence of the current goes to zero showing that anomaly cancellation for the boundary theory is also working as expected. 
Note that, the numerical values for the anomaly ratio $R_{\mathrm{Anomaly}}$ for the right chirality fermion of $Q=5$ is slightly different from the left chirality ones $Q=3,4$. This is expected because the wave functions of the right and the left chirality fermions are not identical on the $s=0$ wall. 

\section{Conclusion} 
In this paper we devised the lattice gradient flow and the equation of motion flow for the slab proposal for chiral gauge theories. While the continuum version of the gradient flow was outlined in \cite{Grabowska:2015qpk}, we uncovered several key subtleties in the explicit lattice construction. The first relates to the fact that the edge states have finite width in the extra dimension, falling off over a length which we denote as $\lambda_\psi/2$ here. Therefore, the flow in the extra dimension must begin when the edge state wave functions have fallen off sufficiently, i.e. beyond the region set by $\lambda_\psi$. Another important point to note is that, the rapid damping of the gauge fields as a function of $|s|$, arising from the flow implies that the bulk Chern-Simons currents far away from the wall are close to zero. Similarly, the anti-wall is decoupled as well with no charge deposition at any time. This is in contrast with the standard domain wall setup where, with the extra dimension independent gauge fields, one expect the wall and the anti-wall to both exhibit charge non-conservation in the presence of nonzero electric field. Of course, the goal of the gauge field flow is to decouple the mirror fermions. Thus it is no surprise that they do not participate in charge or current conservation even when there is charge non-conservation on the wall due to the presence of electric field there. Of course, the higher dimensional theory conserves current exactly and the non-conservation of charge on the wall is compensated by equal and opposite non-conservation of charge in the bulk. 
It is simple to predict these behaviors qualitatively by analytically solving the flow equations and then computing the Chern-Simons current in the bulk. The numerical results confirm the qualitative analytic understanding of the physics. Of course, the natural next step is to generate several two dimensional gauge field configuration, implement gradient flow or EOM flow as outlined in this paper to obtain the bulk gauge fields and then compute the lattice path integral. While, solving the $3-4-5-0$ model in this way would be very interesting, it will also be worthwhile to extend this construction to higher dimension  which should be straightforward. One of the most interesting lattice studies in $n=2$, i.e. for four dimensional domain wall embedded in five dimensional lattice will be to engineer the QCD subsector of the full standard model on the domain wall. In fact, one may simply engineer one-flavor QCD on the domain wall as suggested in \cite{Golterman:2024ccm}, where it was pointed out that the extra dimensional constructions possess an additional exact $U(1)$ symmetry which can turn the $\eta'$ particle into a Goldstone mode, thus conflicting with phenomenology. However, \cite{Kaplan:2024ezz} pointed out that the additional symmetry identified by \cite{Golterman:2024ccm} is consistent with a massive $\eta'$ particle in which case the only remaining effect of the symmetry would be to turn the QCD $\theta$ angle into an unphysical parameter solving the strong CP problem. It will be interesting to examine whether a massive $\eta'$ spectra can be extracted from a lattice path integral of the slab proposal with gauge field flow. Finally, we plan to outline the lattice version of the flow in the disk proposal in future work as well. It will also be interesting to explore the $\eta'$ spectra in that context.   
\section{acknowledgement}
JD acknowledges the support of NSFC under Grants No. 12293060, No. 12293063.
RK is supported in part by the DOE NNSA LRGF Fellowship under cooperative agreement DE-NA0003960.
SS is supported by the U.S. Department of Energy,
Nuclear Physics Quantum Horizons program through the
Early Career Award DE-SC0021892.

\bibliography{ChiRef}

\end{document}